\documentclass[letterpaper, 10 pt, conference]{ieeeconf}  

\IEEEoverridecommandlockouts                              

\usepackage{graphics} 
\usepackage{epsfig} 
\usepackage{mathptmx} 
\usepackage{times} 
\usepackage{amsmath} 
\usepackage{amssymb}  

\usepackage{graphics} 
\usepackage{epsfig} 
\usepackage{mathptmx} 
\usepackage{times} 
\usepackage{amsmath} 
\usepackage{amssymb}  

\usepackage[amsmath,thmmarks]{ntheorem}
\usepackage{algorithmic}
\theorembodyfont{\rmfamily} \theoremstyle{plain}
\newtheorem{theorem}{Theorem}

\newtheorem{definition}{Definition}

\theoremstyle{break}
\newtheorem{algorithm}{Algorithm}

\title{\LARGE \bf
A Method for Crash Prediction and Avoidance \\ Using Hidden Markov Models
}

\author{Avinash Prabu, Lingxi Li, Brian King, and Yaobin Chen 
\thanks{The authors are with the Department of Electrical and Computer Engineering, also with the Transportation Active Safety Institute (TASI), Purdue School of Engineering and Technology, Indiana University-Purdue University Indianapolis (IUPUI), 723 West Michigan Street, SL-160, Indianapolis, IN 46202, USA. Email: 
        {\tt\small \{aprabu, ll7, briking, ychen\}@iupui.edu.}}%
}

\begin{document}

\maketitle
\thispagestyle{empty}
\pagestyle{empty}

\begin{abstract}

In recent years, automotive technology has made a steady progress. In particular, Advanced Driver Assistance System (ADAS) has enabled many safety features in commercial vehicles, for instance, pedestrian detection, lane keeping assist, emergency automatic braking, etc. Although these features provide drivers with a safer operational environment, crashes still happen occasionally due to the complex road conditions and the unpredictable movement of road users including vehicles, pedestrians, bicyclists, and non-motorized vehicles. In this paper, we aim at predicting the possibilities of crashes between vehicles on highway and implementing an appropriate active safety system to prevent the same. In particular, hidden Markov models are developed for the traffic lanes and speed change of vehicles on highway. Algorithms are developed for the prediction of crash probabilities. Simulation experiments are conducted using Matlab, the results illustrate the effectiveness of the proposed research. 
\end{abstract}

\section{INTRODUCTION}
Driving behavior and driver characteristics have been used in the automotive industry to assist the development of active safety systems. A comprehensive study of driving behavior has become one of the main research areas in recent years. Some researchers have also used driving behavior profile as a composite measure of the risk of casualty crash. This paper focuses on using Markov models to predict the driving behavior and calculate the probability of the occurrence of a potential crash thereby taking actions to prevent the same. 

\subsection{Related Work}
The authors in \cite{safetyassessment} suggested a method for assessing the safety of planned trajectories in autonomous vehicles. The future position of the trajectory was computed based on dynamic models and the dynamics of the participants on the road were considered for the prediction of driving behavior. In \cite{drivingintention}, the authors used an artificial neural network (ANN) approach for predicting the maneuvering intentions of a driver to improve the active safety features of the car. The researchers established a hidden Markov model (HMM), which acts as a Bayesian network with two concurrent stochastic process.  The stochastic process was used to describe the maneuvering behaviors of the driver. The ANN was then used to learn the driving conditions and the rules.
In \cite{estimation}, the Markov models were used for the estimation of traffic density in multi-lane roadways. The authors used each lane as a state in the Markov chain and developed a method for the traffic density estimation. 

Other than the works mentioned above, Markov model has also found its significance in learning of driving conditions such as vehicle speed, surrounding traffic speed, and road geometry. In \cite{evolvingmarkovchains}, the authors mainly focused on the energy efficiency in hybrid electric vehicles (HEVs) and adaptive cruise control by capturing the driving conditions using Markov models. The authors used a real-time comparison of Markov chains using Kullback-Liebler (KL) divergence. These were used to characterize each segment of the road with unique characteristics. An on-board learning technique has been used to update the Markov chains.

In \cite{connectedccdesign}, a connected cruise control model was developed using a probabilistic model. The car-following dynamics of the preceding vehicle was then modeled using Markov chains. The connected cruise controller was achieved using a Markov decision-making process. 
In \cite{utilizingmmbc}, model-based communication has been used for V2V communications, based on both small and large scale modeling of the vehicle dynamics. These were coupled into a Markov chain and investigated for implementing cooperative adaptive cruise control. The Markov chain was used for achieving cruise control goals using a predictive method. The authors in \cite{probabilisticanticipation} used a Markov chain predictor on the preceding vehicle behavior. A state space model was created and then a model predictive controller was designed on the state space representation. The predictors were combined with the Gaussian Mixture Model. The analysis showed that the two predictors worked well at different speeds, giving the systems a wide range of operating points.

\subsection{Main Contributions of This Work}
In this paper, we aim at predicting the possibilities of crashes between vehicles on highway and implementing an appropriate active safety system to prevent the same. The main contributions of this paper are summarized as follows.

\begin{enumerate}
    \item Two layers of Markov models for lane and speed change calculation are designed;
    \item An algorithm is developed for the prediction of potential crashes using the developed Markov models;
    \item A methodology is proposed to design the appropriate active safety system, depending on the crash probability;
    \item A Matlab Simulink model is developed to verify the proposed approach.
\end{enumerate}

\section{BASICS OF MARKOV CHAINS}
 Markov chain is a discrete random process and has the memory-less property \cite{c16}. 
 
The general description of a Markov chain is given as follows.
\begin{enumerate}
\item State Space \(\chi\). 
\item Initial state probability vector \(p_0(x)=P[X_0=x]\), \(\forall  x \in \chi \). 
\item Probability of transitions \(p(x, x')\) where \(x\) is the current state and \(x'\) is the future state. 
\end{enumerate}

\subsection{State Transition Matrix}

 Let \(S_i\) be the current state and \(S_j\) be the next state. The state transition probability is given as follows: 
 \begin{equation}
     p_{ij}(k)=P[X_{k+1} = S_j|X_k = S_i]
 \end{equation}
 where \(S_i, \ S_j \in \chi\) and \(k=0, 1, 2,....\)
 It can be observed that the total probability at any state \(S_i\) is 1. 
 \begin{equation}
     \sum_{j}p_{ij}=1.
 \end{equation}
 Note that the process can stay in the same state \(i\) for the next step, its transition probability is denoted by \(p_{ii}\). 
 The transition probabilities in a model are conveniently represented in a matrix form. Let \textbf{P} be the state transition matrix and \(p_{ij}\) be the transition probabilities at its $i$-th row, $j$-th column position. Then we have

\begin{equation}
\label{eq:2}
    \textbf{P}=[p_{ij}] \; \; \; \: \: \:  i,j = 0, 1, 2, ....
\end{equation}

One of the main attributes of a Markov chain is to determine the probabilities of the chain being at a particular state at a specific time instant. These are called state probabilities and are defined as follows, 
\begin{equation}
    \pi_j(k) \equiv P[X_k=j].
\end{equation}

So, the state probability vector at time instant $k$ is defined as
\begin{equation}
    \pi(k) = [\pi_0(k) \ \pi_1(k) \ \pi_2(k) \ \ldots].
\end{equation}

Clearly, the initial state probability vector is defined at the time instant $k=0$ and is given by: 
\begin{equation}
    \pi(0)=[\pi_0(0) \ \pi_1(0) \ \pi_2(0) \ ...].
\end{equation}

\subsection{Ergodic and Regular Markov Chains}

\begin{definition}
An Ergodic Markov chain is a Markov chain in which it is possible to go from every state to every state, not necessarily in one step.
\end{definition}
\begin{definition}
A Regular Markov chain is a Markov chain in which for some power of the state transition matrix, it has only positive elements.
\end{definition}
\begin{theorem}
In a regular Markov chain with a state transition matrix \textbf{P}, as \(n \rightarrow \infty\), \(\textbf{P}^n\) approaches a limiting matrix $W$, which has all rows as the same vector $w$,
\begin{equation}
    \lim_{n \rightarrow \infty} \textbf{P}^n \rightarrow W,
\end{equation}
where the vector $w$ is a probability vector with all elements that are strictly positive and satisfies
\begin{equation}
    w\textbf{P}=w.
    \label{eqn:W}
\end{equation}
\end{theorem}

\begin{theorem}
Given an ergodic Markov chain, its fundamental matrix is defined as \begin{equation}
    Z=(I-\textbf{P}+W)^{-1}
    \label{eqn:Z}
\end{equation}
\noindent If an ergodic Markov chain starts in state $S_i$, the expected number of steps to reach state $S_j$ for the first time is defined as the mean first passage time from $S_i$ to $S_j$ and can be obtained as:

\begin{equation}
    m_{ij}=\dfrac{(Z_{jj}-Z_{ij})}{ w_j},
    \label{eqn:M}
\end{equation}

\end{theorem}
where $Z_{jj}$ is the element in the $j$-th row, $j$-th column position in matrix $Z$, $Z_{ij}$ is the element in the $i$-th row, $j$-th column position in matrix $Z$, and $w_j$ is the $j$-th element in vector $w$.

\section{HIDDEN MARKOV MODELS}
An extension to Markov chains is a complex stochastic process, known as the hidden Markov models (HMMs). In HMMs, the states are not directly visible but the observation sequence, or output, can be observed. The output of the model is directly dependent on the states and thus a pattern for sequence of states can be observed. The term ``hidden" refers to the unknown states in the model. In most practical applications, the states are usually known. The model is called a hidden Markov process even if the states are known \cite{HMM}.

A HMM can be defined using the parameters below. 
\begin{enumerate}
    \item $N$ \(\rightarrow\) the number of states. Even though the states are hidden, in most practical applications, there are relations between observed sequence and the number of states. In some cases the number of states are completely known. 
    \item $M$ \(\rightarrow\) the number of observations at a state. 
    \item \(P=p_{ij}\) \(\rightarrow\) the state transition probability matrix.
    \item \(b_i\) \(\rightarrow\) observation sequence probability at state \(S_i\), where \(i=1,2,3....N\).
    \item \(\pi\) \(\rightarrow\) the initial probability vector.
\end{enumerate}

\section{Design of Active Safety Systems}
Hidden Markov models can be used to design a predictive algorithm to enhance active safety in vehicles. Prediction of driving behavior \cite{driverbehavior} has been one of the main areas of research in active safety systems. The idea here is to design a model that can predict the driving behavior of a particular vehicle and thereby, predicting its location at a particular time. This will help in predicting potential crashes between vehicles by establishing an algorithm that can compare the HMMs of the vehicles, through proper Vehicle-to-Vehicle communications. This information can be used to trigger the appropriate active safety system.
There are some basic assumption to design the HMM, which are summarized below.
\begin{itemize}
    \item All vehicles are equipped with active safety elements. 
    \item All vehicles have capabilities of communicating with each other.
    \item Vehicle dynamics data can be observed and used (including the lane number and vehicle speed).
\end{itemize}

\subsection{Markov Chain - Model Specification}
In this model, two layers of Markov chains are used to develop the prediction flow. The usage of HMM in this paper, slightly deviates from the conventional HMMs. Here, all the states are known and the state transition matrices of both the layers can be computed with the available historical data. 

The first layer of the Markov chain (states) is for lane change. The initial assumption here is that, there are six lanes on each side of the road. The second layer of Markov chain (observed states) consists of speed ranges. The speed probabilities are calculated for each lane to increase the efficiency of the crash prediction. 

\begin{table}
\parbox{.45\linewidth}{
\centering
\begin{tabular}{|c|}
      \hline
      \bf\ States\\
      \hline
      Lane 1\\
     
      Lane 2\\
      
      Lane 3\\
     
      Lane 4\\
      
      Lane 5\\
      
      Lane 6\\
      \hline
      \end{tabular}
\caption{First Layer}
\label{ta:StatesModel1}

}
\hfill
\parbox{.45\linewidth}{
\centering
\begin{tabular}{|c|c|}
      \hline
      \bf\ Observed States &\bf\ Symbol\\
      \hline
      0-10 m/s & a\\
     
      10-20 m/s & b\\
      
      20-30 m/s &c \\
     
      30-40 m/s &d\\
      
      40-50 m/s &e\\
      
      50-60 m/s &f\\
      \hline
    \end{tabular}
\caption{Second Layer}
\label{ta:StatesModel2}
}

\end{table}

\begin{table}[ht]
  \caption{Observation Probabilities for each lane}
  \begin{center}
    \begin{tabular}{|c|c|c|c|c|c|}
      \hline
    \bf 1 & \bf 2 & \bf 3 & \bf 4 & \bf 5 & \bf 6\\
     \hline
      $b_1(a)$ & $b_2(a)$ &$b_3(a)$ & $b_4(a)$ & $b_5(a)$ & $b_6(a)$ \\
      $b_1(b)$ & $b_2(b)$ &$b_3(b)$ & $b_4(b)$ & $b_5(b)$ & $b_6(b)$ \\
      $b_1(c)$ & $b_2(c)$ &$b_3(c)$ & $b_4(c)$ & $b_5(c)$ & $b_6(c)$ \\
      $b_1(d)$ & $b_2(d)$ &$b_3(d)$ & $b_4(d)$ & $b_5(d)$ & $b_6(d)$ \\
      $b_1(e)$ & $b_2(e)$ &$b_3(e)$ & $b_4(e)$ & $b_5(e)$ & $b_6(e)$ \\
      $b_1(f)$ & $b_2(f)$ &$b_3(f)$ & $b_4(f)$ & $b_5(f)$ & $b_6(f)$ \\
       \hline
      
    \end{tabular}
  \end{center}
  \label{ta:ObsProb}
\end{table}

The states of both the layers of the HMM are listed in Tables  \ref{ta:StatesModel1} and \ref{ta:StatesModel2}. The lanes are numbered from one to six from left to right of the road correspondingly. The speed variable has been divided into six states with 10 m/s range. Each observation state is related to a symbol. Table \ref{ta:ObsProb} shows how each observation probability is classified into lanes. 
\subsection{Model Design}
\begin{figure*}[ht]
  \centering
  \includegraphics[scale=0.5] {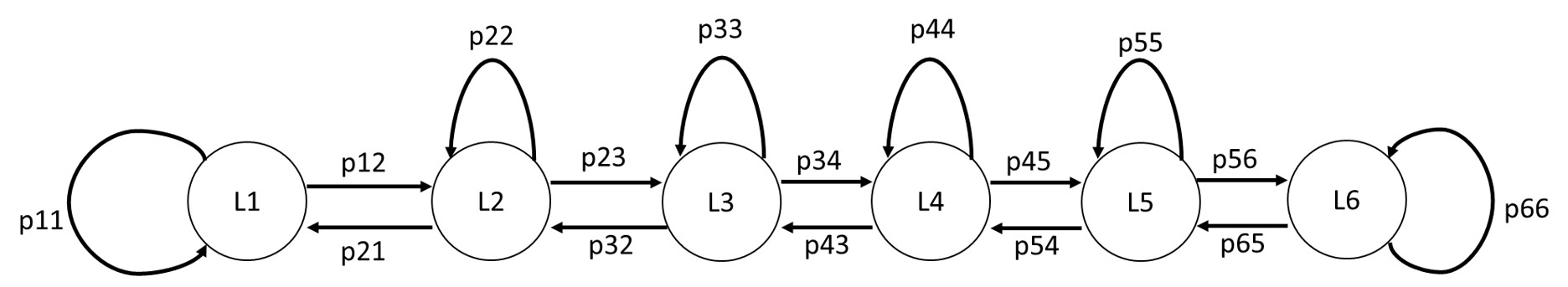}
  \caption{Lane change Markov Chain.}
  \label{fig:Layer1MC}
\end{figure*}

The Markov chain for lane change is depicted in Fig.~\ref{fig:Layer1MC}. L1, L2, L3, L4, L5 and L6 are states discussed above. Practically examining the model, it is not difficult to see that this is an ergodic Markov chain.

\begin{figure*}[ht]
  \centering
  \includegraphics[scale=0.5] {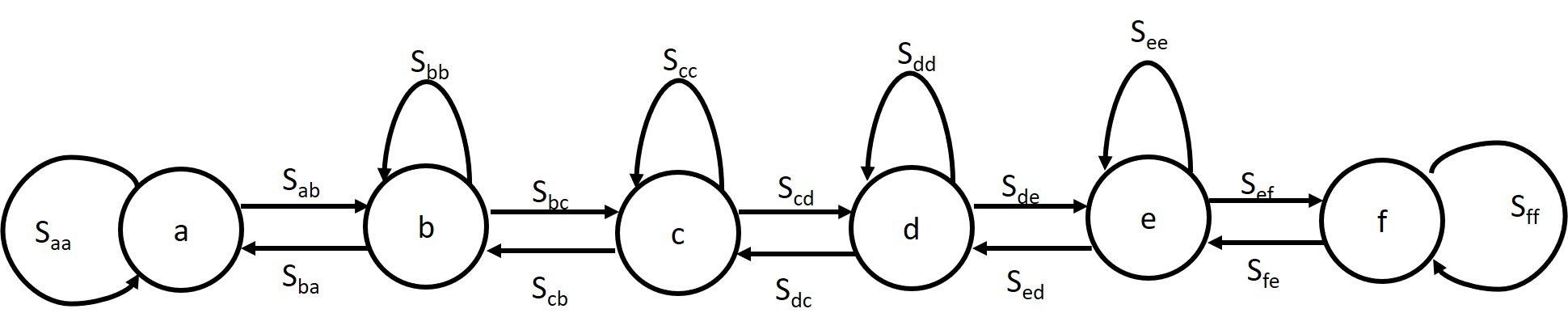}
  \caption{Speed change Markov Chain.}
  \label{fig:Layer2MC}
\end{figure*}

Fig.~\ref{fig:Layer2MC} depicts the second layer of the model, speed change Markov chain. The observed states are a, b, c, d, e, and f which correspond to the speed ranges. 

\subsubsection{Lane change state transition matrix, P}
The state transition matrix is obtained from historical lane data of a vehicle. The formula used for calculating each element of the state transition matrix is discussed in equation (\ref{eqn:P}). 

\begin{multline}
 \text{Probability from $i$ to $j$}\\
 =\dfrac{\text{No. of rows with current lane =$i$, and next lane =$j$}}{\text{No. of rows with current lane =$i$}}
\end{multline}

\begin{equation}
    p_{ij}=\dfrac{\sum_{k=1}^n\eta(X_k=i|X_{k+1}=j)}{\eta(X_k=i)}
    \label{eqn:P}
\end{equation}

\begin{equation}
    P=\{p_{ij}\}
\end{equation}

where, 
\(i, j = \{1,2,3,4,5,6\}\)

$i \rightarrow$ current lane, $j \rightarrow$ next lane

$\eta \rightarrow$ number of occurrences

$n \rightarrow$ total number of data rows

\subsubsection{Speed change state transition matrix, S}
The speed change data is also obtained from historical data. The formula for calculating the same is given in equation (\ref{eqn:S}). 

\begin{equation}
    s_{(l-m)-(m-n)}=\dfrac{\sum_{k=1}^n\eta(l<Y_k<m|m<Y_{k+1}<n)}{\eta(l<Y_k<m)}
    \label{eqn:S}
\end{equation}

\begin{equation}
    S=\{s_{(l-m)-(m-n)}\}
\end{equation}

where, 

($l$ to $m$), ($m$ to $n$) = \(\{a,b,c,d,e\}\)

$\eta \rightarrow$ number of occurrences

$n \rightarrow$ total number of data rows

\subsubsection{Limiting matrix (W), Fundamental matrix (Z) and Mean first passage matrix (M)}
The mean first passage matrix finds its use in determining which active safety system needs to be triggered. The formula for calculating $W$, $Z$, and $M$ can be found in equations (\ref{eqn:W}), (\ref{eqn:Z}), and (\ref{eqn:M}).

 \subsubsection{Probability of position at time=$t$}
 The probability of the vehicle being at a particular lane can be found using the initial probability vector and the state transition matrix. The formula is given in equation (\ref{eqn:pi}).
 
 \begin{equation}
     \pi_t=\pi_0P^t
     \label{eqn:pi}
 \end{equation}
 $\pi_t$ has six elements, each of which gives the lane probability at time $t$. 
 
 \section{Crash Prediction and Avoidance}
 The prediction of crash has three parts, each of which has been explained using flow charts in the upcoming subsection.
 \subsection{Flow-1 - Calculating the probable time of crash}
 
 \begin{algorithm}
 \vspace{0.1in}
 \noindent START: Compute state transition matrix of Car 1 ($P1$), and Car 2 ($P2$)
 
 \noindent 1. Determine the current lane number of both the vehicles. 
 
 $i$ $\rightarrow$ Car 1, 
 $j$ $\rightarrow$ Car 2. 
 
 This is used to determine the initial probability vector of both the vehicles. 
 
  \noindent 2. Determine the current speed of both the vehicles.
  
 $V1$ $\rightarrow$ Car 1, 
 $V2$ $\rightarrow$ Car 2
 
 \noindent 3. Calculate relative velocity ($V$)
 
 V=V2-V1
 
  \noindent 4. \begin{algorithmic}
 \IF {V $>$ 0}
        \STATE GOTO 5
\ELSE 
        \STATE GOTO START
\ENDIF
\end{algorithmic}
 
 \noindent 5. Obtain horizontal distance between the cars (d). Refer to Fig.~\ref{fig:lidar}, equations (\ref{eqn:Lidar}) and (\ref{eqn:Lidar2}). 
 
 \noindent 6. Determine the time at which the cars will be on the same y-coordinate (t). It is depicted in equation (\ref{eqn:t}). 
 
 \noindent 7. Determine the speed change probabilities of both the vehicles at respective lanes. 
 
 \noindent 8.
 \begin{algorithmic}
  \IF{change in speed probability greater than 0.5}
        
        \STATE GOTO 2
  
  \ELSE
        \STATE GOTO Next Flow
    \ENDIF
        
 \end{algorithmic} 

 $\hfill$ $\Box$
 \end{algorithm}
 
 A Light detection and ranging (Lidar) device is used to determine the y-distance between the cars. The technique used is summarized below. 
 
 \subsubsection{Lidar operation}
\begin{figure}[ht]
  \centering
  \includegraphics[scale=0.5] {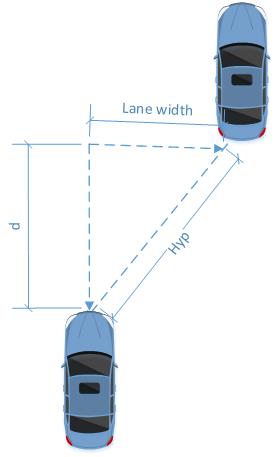}
  \caption{Lidar principle.}
  \label{fig:lidar}
 \end{figure}

 A Lidar uses pulsed laser to determine the distance between two vehicles. Lidar has a 360 degree visibility with an extremely accurate distance measurement. Once the Lidar sends a pulse of laser, it determines the time it takes for the light to return back to the sensor. This concept is illustrated in Fig.~\ref{fig:lidar}. The diagonal distance between the cars are determined by the Lidar. The formula embedded in the device is presented in equation (\ref{eqn:Lidar}).  The equation to calculate the y-distance (d) is explained in equation (\ref{eqn:Lidar2}).  
 
 \begin{equation}
     Hyp=\dfrac{(Ltime)(c)}{2}
     \label{eqn:Lidar}
 \end{equation}
 
 $Ltime$ is the time elapsed by the Lidar to return back.

 \begin{equation}
     d=\sqrt{Hyp^2-LW^2}
     \label{eqn:Lidar2}
 \end{equation}
 
 \subsubsection{Probable time of crash}
The probable time $t$, is calculated using the formula in equation (\ref{eqn:t}). 
\begin{equation}
    t=d/V
    \label{eqn:t}
\end{equation}
 
Now that the probable time of the crash has been calculated. To verify that the cars will maintain the same speed, the state transition matrix for speed change is calculated. If the probability of speed change is higher than 0.5, then the loop goes back to determine the current speed. Otherwise, the flow goes into the next step. 

\subsection{Flow 2 - Calculating the probabilities of potential crashes}
\begin{algorithm}
 \vspace{0.1in}
  \noindent 1. Calculate the state transition matrix of Car 1 and Car 2 at time=$t$ ($P_1^t$ and $P_2^t$)
  
  \noindent 2. Calculate the probability of position (Lane) of both the cars at time=$t$
 
$\pi1_t = (\pi1_0)(P1^t), \; \text{Car 1}$
 
$\pi2_t = (\pi2_0)(P2^t), \; \text{Car 2}$
 
 \noindent 3.  Determine the probability of both the cars being at the same lane by performing element-wise multiplication of  $\pi1_t$ and $\pi2_t$ 
 
 \begin{algorithmic} 
 \FOR {k=1 \TO 6}
         \STATE $PC(k)= (\pi1_t(k))(\pi2_t(k))$
 \ENDFOR
 \end{algorithmic}
 
 \noindent 4. GOTO Next Flow
 $\hfill$ $\Box$
 \end{algorithm}
 
 $\pi1_t$ and $\pi2_t$ are probability vectors, at time t, which provide the probability of the cars being at lanes, 1 to 6, respectively. $PC$ is also a probability vector which provides the probability of both the cars being at the same lane at time, t. $PC$ here is obtained through element-wise multiplication of $\pi1_t$ and $\pi2_t$. This vector provides the potential crash probabilities at every lane at time, t. 
 
 \subsection{Flow 3 - Determining the appropriate active safety systems}
  \begin{algorithm}
 \vspace{0.1in}
 \noindent INPUT: Probable crash probabilities $PC$ and Distance between the vehicles (to determine the car in front) 
\vspace{0.1in}

\begin{algorithmic}
 \FOR{k=1 \TO 6}
 
    \IF{PC(k) $\geq$ 0.3}
        
        \STATE Compute limiting and fundamental matrices.
        \STATE Compute mean first passage matrix for Car1 (M1) and Car2 (M2). 
        \vspace{0.1in}
        \IF{M1(i,k) $>$ t}
            
            \IF{Car1 is in front}
                
                \STATE Adaptive cruise control ON for Car2
            
            \ELSIF{Car2 is in front}
                
                \STATE Adaptive cruise control ON for Car1
                
            \ENDIF
        \vspace{0.1in}
        \ELSIF{M1(j,k) $>$ t} 
            
            \IF{Car1 is in front}
                
                \STATE Adaptive cruise control ON for Car2
            
            \ELSIF{Car2 is in front}
                
                \STATE Adaptive cruise control ON for Car1
                
            \ENDIF
        \vspace{0.1in}
        \ELSE
            
            \STATE Lane departure and steering on for Car1 and Car2
        
        \ENDIF
    \vspace{0.1in}
    \ENDIF
    
\ENDFOR

\end{algorithmic}
 
 $\hfill$ $\Box$
 \end{algorithm}
 
 The first step is to determine whether the calculated probability of crash at each lane is at least one-third. The mean first passage matrix helps in determining whether the car changes the lane before time $t$. We assume that $i$ and $j$ are still the current lanes for Car1 and Car2 respectively. If the $(i,k)th$ element of M1 or $(j,k)th$ element of M2 is higher than $t$, it will lead to a rear end crash. Thus, adaptive cruise control is ON for the following car. If the $(i,k)th$ element of M1 or $(j,k)th$ element of M2 is less than $t$, it means there is a possibility of sideways collision. Thus the lane departure and steering assist is ON for both the vehicles. 
 
\section{SIMULATION AND RESULTS}
The simulation of the whole system was carried out in MATLAB/SIMULINK. The database used for simulation was obtained from U.S. Department of Transportation's (US-DOT) Intelligent Transportation Systems (ITS) \cite{originator}. The database of two cars were obtained and simulated into different scenarios to validate the proposed methods. 

Software programs were developed to obtain lane change state transition matrix as explained in equation (\ref{eqn:P}). It is observed that the sum of elements in each row of the state transition matrix is equal to 1. The state transition matrix also gives us an idea of how often a driver changes lane. The speed change matrix evaluation follows the same algorithm as in the lane change matrix. Matlab's MPC toolbox is used to determine the acceleration/deceleration required for the ego vehicle to maintain a safe distance from the target vehicle. The two main objectives of the ACC block are explained below: 
\begin{enumerate}
    \item If the relative distance is greater than the safe distance, the ego vehicle travels in the speed set by the driver.
    \item If the relative distance is less than the safe distance, the control goal is to reduce the speed of the ego vehicle so that the distance between the cars is maintained at a safe distance. 
\end{enumerate}

\subsection{Simulation Scenarios and Results}
\subsubsection{Scenario 1 -- Unsafe operation with ACC}
\begin{itemize}
    \item There are six lanes on one side of the road.
    \item Car 1 is at lane 6, traveling in front of Car 2 at 30 $m/s$ with a constant acceleration of 0.6 $m/s^2$.
    \item Car 2 is at lane 5, traveling at 40$m/s$.
    \item The relative distance between the cars is 40m.
\end{itemize}

\begin{equation}
    \pi_1=
    \begin{bmatrix}
    0&0&0&0&0&1
    \end{bmatrix},
    \label{eqn:pi1}
\end{equation}
\begin{equation}
    \pi_2=
    \begin{bmatrix}
    0&0&0&0&1&0
    \end{bmatrix},
    \label{eqn:pi2}
\end{equation}
\begin{equation*}
t=40/(40-30)=4.
\end{equation*}
\begin{equation*}
    \pi1_4(5)=0.86469, \; \pi2_4(5)=0.88792.
\end{equation*}
\begin{equation*}
    \text{Crash Probability at lane 5}=(0.86469)(0.88792)=0.7678.
\end{equation*}

From the matrices obtained from the Matlab program, it was clear that Car 1 and Car 2 have a higher probability to be at lane 5 at the probable time of crash. Comparing the probable time of crash with \(M1(6,5)=1.3s\), it is possible that Car 1 changes its lane much earlier than the time of crash. So this indicates that ACC for Car 2 has to be turned ON to prevent the rear-end crash. 

\subsubsection{Scenario 2 -- Unsafe operation with lane departure and steering assist}

\begin{itemize}
    \item There are six lanes on one side of the road.
    \item Car 1 is at lane 6, traveling at 50 $m/s$ with a constant acceleration of 0.6 $m/s^2$.
    \item Car 2 is at lane 5, traveling in front of car 1 at 40$m/s$.
    \item The relative distance between the cars is 12m.
\end{itemize}

The initial probability vector stays the same for both vehicles as in Scenario 1. 
\begin{equation*}
t=12/(50-40)=1.2.
\end{equation*}
\begin{equation*}
    \pi1_1.2(5)=0.0.7849, \; \pi2_1.2(5)=0.96022.
\end{equation*}
\begin{equation*}
    \text{Crash Probability at lane 5}=(0.7849)(0.96022)=0.7536.
\end{equation*}
It can be seen from the results that there is a high probability that the cars will be at lane 5 at the same time. The time of crash is compared with the mean first passage time of Car 1 from lane 6 to lane 5 and if it is lesser, then lane departure warning and steering assist are ON for Car 1, which restrict Car 1 going from lane 6 to lane 5 and the potential crash is avoided.

\subsubsection{Scenario 3 -- Safe operation}
\begin{enumerate}
    \item There are six lanes on one side of the road.
    \item Car 1 is at lane 1, traveling in front of Car 2 at 50~$m/s$.
    \item Car 2 is at lane 2, traveling at 60~$m/s$.
    \item The relative distance between the cars is 30m.
\end{enumerate}
\begin{equation*}
t=30/(60-50)=3.
\end{equation*}
\begin{equation*}
    \pi1_3(2)=0.043, \; \pi2_3(1)=0.052.
\end{equation*}

It can be seen that the probability of crash between two cars is really small. Thus, they do not need to change lanes at the probable time of crash. 

\section{Conclusion}

In this paper, we developed two layers of HMMs for vehicle dynamics on highway, one for lane change and one for speed change. The model parameters were obtained from the USDOT traffic data. Algorithms were developed for the prediction of potential crashes using the developed models and methodologies were proposed to design the appropriate active safety system, depending on the crash probability. Simulation experiments were also carried out and several scenarios were presented to illustrate the effectiveness of the proposed approach.

One future research direction is to consider databases that include more parameters on driver behavior. It is also interesting to investigate more advanced models for crash prediction and prevention.






\end{document}